\documentclass[aps,twocolumn,groupedaddress,superscriptaddress,nofootinbib,amsmath,amssymb]{revtex4}
\usepackage{amsbsy,subfigure,hyperref,bbm,times}
\usepackage[T1]{fontenc}
\usepackage{epsfig}
\usepackage{color}
\usepackage{graphicx}
\usepackage{dcolumn}
\usepackage{bm}
\newcommand{\bra}[1]{\langle#1|}
\newcommand{\ket}[1]{|#1\rangle}
\newcommand{\ave}[1]{\langle#1\rangle}

\newcommand{\scal}[2]{\langle#1|#2\rangle}
\providecommand{\openone}{\leavevmode\hbox{\small1\kern-3.8pt\normalsize1}}
\newcommand{\sbt}{\,\begin{picture}(0.3,0.3)(-1,-2) \circle*{2}\end{picture}\ }

\begin{document}
\title{Quantum entanglement of identical particles by standard information-theoretic notions}
\author{Rosario Lo Franco}
\email{rosario.lofranco@unipa.it}
\affiliation{Dipartimento di Energia, Ingegneria dell'Informazione e Modelli Matematici, Universit\`{a} di Palermo, Viale delle Scienze, Ed. 9, 90128 Palermo, Italy}
\author{Giuseppe Compagno}
\affiliation{Dipartimento di Fisica e Chimica, Universit\`a di Palermo, via Archirafi 36, 90123 Palermo,
Italy}

\date{\today }

\begin{abstract}
Quantum entanglement of identical particles is essential in quantum information theory. Yet, its correct determination remains an open issue hindering the general understanding and exploitation of many-particle systems. Operator-based methods have been developed that attempt to overcome the issue. 
We introduce a state-based method which, as second quantization, does not label identical particles and presents conceptual and technical advances compared to the previous ones. It establishes the quantitative role played by arbitrary wave function overlaps, local measurements and particle nature (bosons or fermions) in assessing entanglement by notions commonly used in quantum information theory for distinguishable particles, like partial trace. Our approach furthermore shows that bringing identical particles into the same spatial location functions as an entangling gate, providing fundamental theoretical support to recent experimental observations with ultracold atoms. These results pave the way to set and interpret experiments for utilizing quantum correlations in realistic scenarios where overlap of particles can count, as in Bose-Einstein condensates, quantum dots and biological molecular aggregates.  
\end{abstract}


\maketitle

Entanglement of identical particles is fundamental in understanding and exploiting composite quantum systems \cite{horodecki2009RMP,amico2008RMP}, being a resource for scalable quantum information tasks \cite{amico2008RMP,giovannetti2004Science,riedel2010Nature,benatti2014NJP,cramer2013NatComm,marzolino2015}, for instance in Bose-Einstein condensates (ultracold gases in an optical lattice) \cite{bloch2008RMP,hayes2007PRL,anderlini2007Nature} or in quantum dots \cite{petta2005Science,lundskog2014,tan2015PRL,veldhorst2015Nature}. Therefore, its correct determination becomes a central requirement in quantum information theory. However, in systems of identical particles a new aspect emerges compared to systems of distinguishable particles, namely the role played by quantum particle indistinguishability in entanglement determination, which remains debated \cite{Li2001PRA,Paskauskas2001PRA,cirac2001PRA,eckert2002AnnPhys,plastino2009EPL,ghirardi2004PRA,mintertreview,ghirardi2002JSP,tichyFort,buscemi2007PRA,vogel2015PRA,Shi2003PRA,giulianoEPJD,balachandran2013PRL,benatti2012AnnPhys,sasaki2011PRA,benatti2012PRA,benatti2014review}. For instance, there is to date no general agreement either on the very simple case if two identical particles in the same site are entangled or not \cite{cirac2001PRA,ghirardi2004PRA,mintertreview,ghirardi2002JSP,tichyFort,cavalcanti2007PRB,plenio2014PRL}, although there are recent experiments where this situation is analyzed \cite{tan2015PRL,kaufman2015Nature}. 

The main issue is that usual entanglement measures of quantum information theory, such as the von Neumann entropy of the reduced state, fail to be directly applied to identical particle states because they witness entanglement even for independent separated particles which are clearly uncorrelated, also showing contradictory results for bosons and fermions \cite{ghirardi2004PRA,mintertreview}. We remark that this issue exists not only in the particle-based (first quantization) description \cite{ghirardi2004PRA,mintertreview} but also in the mode-based (second quantization) one \cite{Li2001PRA,Paskauskas2001PRA}, where name labels do not explicitly appear but are however implicitly assumed. 
This problem has induced to develop methods to identify identical particle entanglement that are at variance with respect to the usual ones adopted for nonidentical particles, by redefining the notion of entanglement \cite{cirac2001PRA,eckert2002AnnPhys,plastino2009EPL,ghirardi2002JSP,ghirardi2004PRA,mintertreview,tichyFort,buscemi2007PRA,vogel2015PRA,Shi2003PRA,giulianoEPJD} or searching for tensor product structures supported by the observables \cite{balachandran2013PRL,benatti2012AnnPhys,sasaki2011PRA,benatti2012PRA,friis2013PRA,benatti2014review}, and whose aim is to discriminate the physical part of entanglement from the unphysical one. This necessity of new notions to discuss quantum correlations for identical and nonidentical particles looks surprising. Moreover, these approaches remain somewhat technically awkward and not well suited to quantify entanglement under general conditions of scalability and realistic scenarios where the constituting identical particles are close together to spatially overlap \cite{bloch2008RMP,hayes2007PRL,anderlini2007Nature,tan2015PRL,veldhorst2015Nature,jesenkoNJP}. These drawbacks jeopardize analysis of entanglement and interpretation of experiments both under complete overlap and more strongly in the case of partial overlap. Thus, the relationship between entanglement and identity of particles is still an open issue from both conceptual and practical viewpoint hindering the general understanding and exploitation of composite quantum systems made of identical particles. 

In quantum mechanics, name-labels are assigned to identical particles making them distinguishable. In order that this new fictitious system behaves as the real bosonic or fermionic one only the symmetrized or antisymmetrized states with respect to labels are permitted \cite{cohenbook,feynmanquantum}. While this procedure works well in the usual practice, when it comes to entanglement which crucially depends on the form of the state vector, confusion arises linked to appearance of simultaneous real and fictitious (label born) contributions to it.  

In this work we aim at providing an advancement towards the straightforward description of quantum correlations in identical particle systems grounded on simple physical arguments which can unambiguously answer the general question: when and at which extent quantum particle indistinguishability assumes physical relevance in determining the entanglement among the particles? 
We present here a treatment of identical particles which, like the second quantization, does not resort to name labels yet adopting a particle-based (first quantization) formalism in terms of states. This approach assumes that a many-particle state is a whole single object, characterized by a complete set of commuting observable, and quantifies the physical entanglement of bosons and fermions on the same footing by the same notions used for distinguishable particles such as the von Neumann entropy of partial trace. It allows the study of identical particle entanglement under arbitrary conditions of wave function overlap at the same complexity level required for nonidentical particles and, albeit presented here for two particles, it is straightforwardly generalizable to many-particle systems for scalability. The known results for distinguishable particles can be also retrieved by imposing the condition of spatially separated (i.e., non overlapping) particles. Our approach quantitatively establishes the role of local measurements, particle nature and spatial overlap in assessing identical particle entanglement, supplying theoretical support to very recent experimental observations of entangling operations for identical atoms \cite{kaufman2015Nature}.

\section{Results}

\subsection{Description of the new approach} 
\begin{figure}[t!]
\begin{center}
{\includegraphics[width= 0.46\textwidth]{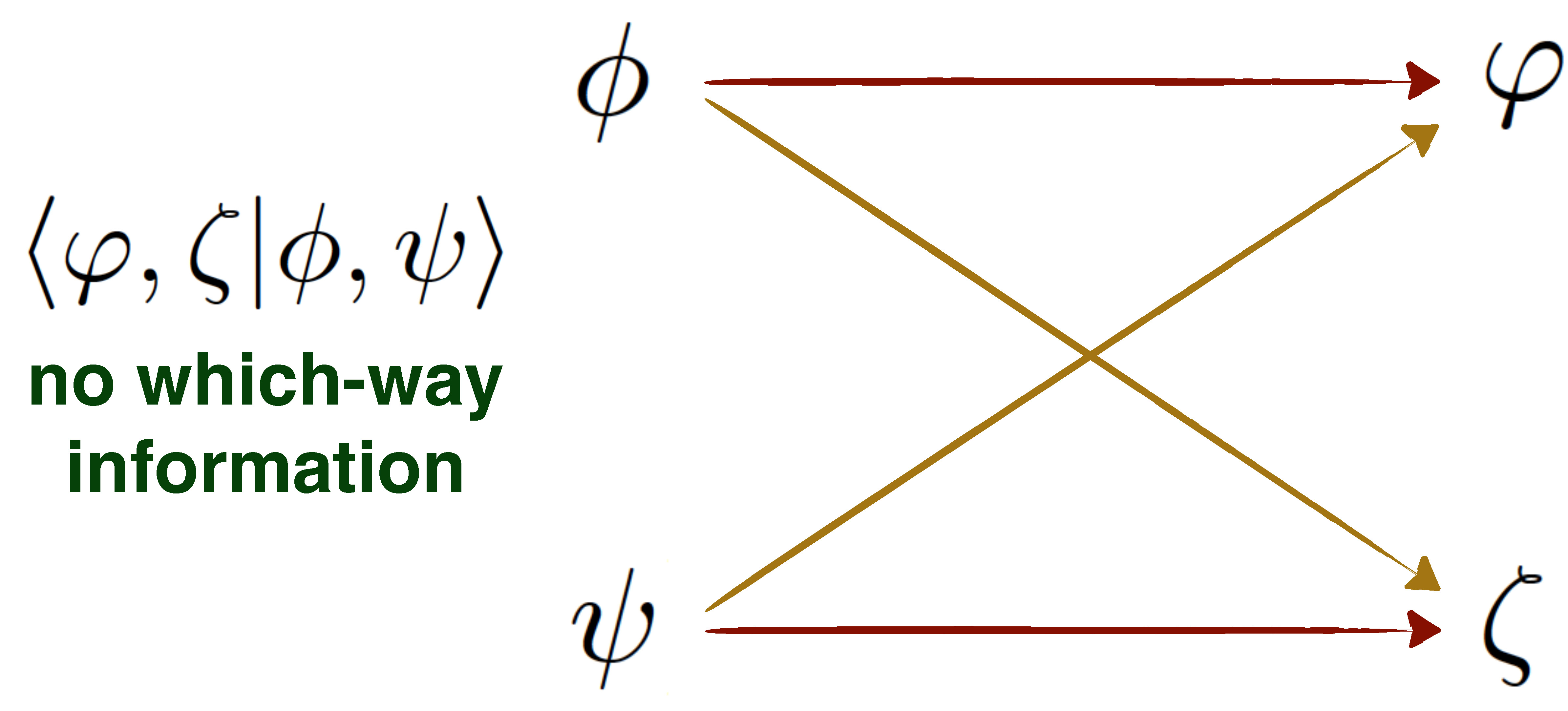}}
\end{center}
\caption{The probability amplitude $\scal{\varphi,\zeta}{\phi,\psi}$ of Eq. (\ref{scalarproduct}) originates quantum mechanically from the lack of which-way information: the transition of one particle to $\varphi$ ($\zeta$) can equally come from $\phi$ and $\psi$.} 
\label{fig:nowhichway}
\end{figure}

Indistinguishability requires that the identical particles cannot be individually addressed and, in accordance to quantum mechanics, introduction of unphysical quantities to treat them is thus unneeded. 
In fact, the system can be completely described in terms of observables determining the one-particle states. 
The global state, taken as the set of one-particle states, must be considered as a ``holistic'' indivisible entity.  
We illustrate this point by taking for simplicity a system of two identical particles whose state vector describes one particle in the state $\phi$ and one in $\psi$: it is thus completely characterized by enumerating the states and represented as $\ket{\phi,\psi}$. The physical predictions on the system follow from the two-particle probability amplitudes $\scal{\varphi,\zeta}{\phi,\psi}$, where $\varphi$, $\zeta$ are one-particle states of another global two-particle state vector. 
We assume that it can be expressed by means of the one-particle amplitudes $\scal{l}{r}$ ($l=\varphi,\zeta$; $r=\phi,\psi$) and linearly depends on one-particle states. Due to the indistinguishability, the probability amplitude of finding one particle in $\varphi$ ($\zeta$) comes from having one particle in $\phi$ or $\psi$, as illustrated in Fig.~\ref{fig:nowhichway}. By simply applying the quantum mechanical superposition principle of alternative paths \cite{feynmanquantum}, we define the two-particle probability amplitude $\scal{\varphi,\zeta}{\phi,\psi}$ as a symmetrized inner product of two state vectors in terms of a linear combination with same weight of products of one-particle amplitudes
\begin{equation}\label{scalarproduct}
\scal{\varphi,\zeta}{\phi,\psi}:= \scal{\varphi}{\phi}\scal{\zeta}{\psi}+\eta\scal{\varphi}{\psi}\scal{\zeta}{\phi}, 
\end{equation} 
where $\eta^2=1$. This equation constitutes the core of our approach and directly encompasses the required particle spin-statistics symmetry, or symmetrization postulate \cite{cohenbook,goyal2014} (see appendix \ref{exchange} about particle exchange in our approach). 
The right-hand side of Eq.~(\ref{scalarproduct}) induces a symmetry with respect to the swapping of one-particle state position within the two-particle state vector: $\scal{\varphi,\zeta}{\phi,\psi}=\eta\scal{\varphi,\zeta}{\psi,\phi}$ or $\ket{\phi,\psi}=\eta\ket{\psi,\phi}$. 
The probability amplitude of finding the two particles in the same state $\varphi$ is $\scal{\varphi,\varphi}{\phi,\psi}=(1+\eta)\scal{\varphi}{\phi}\scal{\varphi}{\psi}$.
As usual, according to the Pauli exclusion principle this must take the minimum value (zero) for fermions which gives $\eta=-1$, while the maximum value for bosons implying $\eta=+1$. 
Linearity of the two-particle state vector with respect to one-particle states immediately follows from the linearity of the one-particle amplitudes in Eq.~(\ref{scalarproduct}). The two-particle state vectors thus span a no-label symmetric state space $\mathcal{H}^{(2)}_\eta$.
In general, as in second quantization, a two-particle state $\ket{\widetilde{\Phi}}=\ket{\varphi_1,\varphi_2}$ is not normalized. The corresponding normalized state $\ket{\Phi}$, such that $\scal{\Phi}{\Phi}=1$, is $\ket{\Phi}=(1/\sqrt{\mathcal{N}})\ket{\varphi_1,\varphi_2}$, where $\mathcal{N}=1+\eta|\scal{\varphi_1}{\varphi_2}|^2$ is obtained by Eq.~(\ref{scalarproduct}). For orthogonal one-particle states, $\scal{\varphi_1}{\varphi_2}=0$, one has $\mathcal{N}=1$ and 
$\ket{\Phi}=\ket{\varphi_1,\varphi_2}$.

A one-particle operator $A^{(1)}$, according to the standard definition \cite{cohenbook}, acts on a two-particle state as $A^{(1)}\ket{\varphi_1,\varphi_2}:= \ket{A^{(1)}\varphi_1,\varphi_2}+\ket{\varphi_1,A^{(1)}\varphi_2}$. Its expectation value on a normalized state is $\ave{A^{(1)}}_{\Phi}:= \bra{\Phi}A^{(1)}\ket{\Phi}$. In a one-particle state space with basis $\mathcal{B}^{(1)}=\{\ket{\psi_k},k=1,2,\ldots\}$, in general $A^{(1)}=\sum_{j,k}a_{jk}\ket{\psi_j}\bra{\psi_k}$. Using symmetry and linearity from Eq.~(\ref{scalarproduct}), it is straightforward to show that $\ket{\psi_j}\bra{\psi_k}\sbt\ket{\varphi_1,\varphi_2}=\ket{\psi_j,\scal{\psi_k}{\varphi_1}\varphi_2+\eta \scal{\psi_k}{\varphi_2}\varphi_1}$. 
We now define a non-separable symmetric external product of one-particle states 
$\ket{\varphi_1,\varphi_2}:=\ket{\varphi_1}\times \ket{\varphi_2}$, from which $\bra{\varphi_1,\varphi_2}=(\ket{\varphi_1}\times \ket{\varphi_2})^\dag=\bra{\varphi_2}\times \bra{\varphi_1}$ and $\ket{\varphi_1}\times \ket{\varphi_2}=\eta\ket{\varphi_2}\times \ket{\varphi_1}$ (for fermions it recalls Penrose's wedge product defined in terms of labelled states \cite{penrosebook}). This permits us to write $\ket{\psi_j}\bra{\psi_k}\ket{\varphi_1,\varphi_2}=\ket{\psi_j}\times(\scal{\psi_k}{\varphi_1}\ket{\varphi_2}+\eta \scal{\psi_k}{\varphi_2}\ket{\varphi_1})$ which defines a symmetric inner product between state spaces of different dimensionality
\begin{equation}\label{reducedpurestate}
\bra{\psi_k}\sbt\ \ket{\varphi_1,\varphi_2}\equiv
\scal{\psi_k}{\varphi_1,\varphi_2}=\scal{\psi_k}{\varphi_1}\ket{\varphi_2}+\eta \scal{\psi_k}{\varphi_2}\ket{\varphi_1}.
\end{equation}
This equation provides the unnormalized reduced one-particle pure state obtained after projecting a two-particle state on $\ket{\psi_k}$ (one-particle projective measurement).
Consider now the one-particle projection operator $\Pi_k^{(1)}=\ket{\psi_k}\bra{\psi_k}$ and then define the one-particle identity operator $\mathbb{I}^{(1)}=\sum_{k} \Pi_k^{(1)}$, such that $\mathbb{I}^{(1)}\ket{\varphi}=\ket{\varphi}$. It is immediate to see that $\mathbb{I}^{(1)}\ket{\Phi}=\mathbb{I}^{(1)}\ket{\varphi_1,\varphi_2}/\sqrt{\mathcal{N}}=2\ket{\Phi}$ and thus $\ave{\mathbb{I}^{(1)}}_{\Phi}=2$. Therefore, the normalized reduced one-particle pure state $\ket{\phi_k}$ and the probability $p_k$ to observe it after the projective measurement $\Pi_k^{(1)}$ are, respectively,
\begin{equation}\label{projection}
\ket{\phi_k}=\scal{\psi_k}{\Phi}/\sqrt{\ave{\Pi_k^{(1)}}_{\Phi}},
\quad p_k = \ave{\Pi_k^{(1)}}_{\Phi}/2,
\end{equation}     
where $\scal{\psi_k}{\Phi}=\scal{\psi_k}{\varphi_1,\varphi_2}/\sqrt{\mathcal{N}}$ is obtained by Eq.~(\ref{reducedpurestate}) and $\sum_k p_k=1$.  
If the two one-particle states are orthonormal ($\scal{\varphi_1}{\varphi_2}=0$) then $p_k=(|\scal{\psi_k}{\varphi_1}|^2+|\scal{\psi_k}{\varphi_2}|^2)/2$, which corresponds to the sum of probabilities of two incompatible outcomes, as expected. 
The partial trace of a system is physically interpreted as the statistical ensemble of all the normalized reduced states obtained after projective measurement on the basis states, that operationally corresponds to measure a subsystem particle without registering the outcomes \cite{horodecki2009RMP,amico2008RMP}. Hence, from Eq.~(\ref{projection}) we immediately determine the one-particle reduced density matrix as    
\begin{equation}\label{partialtrace}
\rho^{(1)} = \sum_{k} p_k \ket{\phi_k}\bra{\phi_k} = (1/\mathcal{\widetilde{N}})\ \mathrm{Tr}^{(1)}\ket{\widetilde{\Phi}}\bra{\widetilde{\Phi}},
\end{equation}  
where $\mathrm{Tr}^{(1)}\ket{\widetilde{\Phi}}\bra{\widetilde{\Phi}}=\sum_{k}\scal{\psi_k}{\varphi_1,\varphi_2}\scal{\varphi_1,\varphi_2}{\psi_k}$ 
and $\mathcal{\widetilde{N}}=2\mathcal{N}$. For calculation convenience, we emphasize that $\rho^{(1)}$ is obtained starting from unnormalized two-particle states and finally introducing a normalization constant $\mathcal{\widetilde{N}}$ such that $\mathrm{Tr}^{(1)}\rho^{(1)}=1$. We stress that the definition of partial trace given above is a physical operation on the system state, based on effective projective measurements, which never suffers the controversies exhibited by the unphysical partial trace operation performed in the description with name labels (see appendix \ref{comparison} for comparison and discussions) \cite{ghirardi2004PRA,mintertreview}.  

As a consequence of remaining within $\mathcal{H}^{(2)}_\eta$, we can exploit the ordinary notion that the degree of mixing of the reduced density matrix is directly related to the amount of entanglement of the global pure state \cite{horodecki2009RMP}. 
Entanglement is a nonlocal quantum feature and its presence in composite systems of nonidentical particles is individuated by local measurements made on the individual particles \cite{horodecki2009RMP,amico2008RMP,mintertreview}.
To quantify entanglement of identical particles, which are individually unaddressable, a suitable definition of one-particle measurement must be given which requires the condition of locality. Here we give the following:

\textbf{Definition.} \emph{A local one-particle measurement for systems of identical particles is the measurement of a property of one particle performed on a localized region of space $M$ (site or spatial mode) where the particle has nonzero probability of being found}. 

We stress that this definition is in perfect analogy with the meaning of local measurement in ordinary Bell nonlocality tests  \cite{horodecki2009RMP,amico2008RMP,ghirardi2004PRA} and, in the case of spatially separated particles, reduces to the usual measurement on a given (addressable) particle. In fact, identical particles are recognized behaving as nonidentical when they live in spatially separated modes \cite{facchiarxiv}, that is a natural request in recovering the distinguishability of bosons and fermions in experiments \cite{herbut2001AJP}. 
It is known that nonlocal measurements on uncorrelated spatially separated identical particles produce the so-called ``measurement-induced entanglement'' \cite{mintertreview,roch2014PRL} (see appendix \ref{measurement-induced}). 
Along this work, we are only interested to the entanglement determined by local measurements according to the above definition. 
Peculiar particle identity effects on entanglement are expected to manifest when particles are close enough to have overlapping spatial modes. 

Due to these fundamental preliminary results of our new approach, the entanglement $E(\Phi)$ of a pure state of two identical particles can be then quantified via the Von Neumann entropy of the one-particle reduced density matrix derived by the \emph{localized partial trace}, which is obtained by Eq.~(\ref{partialtrace}) with the sum over the index $k$ limited to the subset $k_M$ corresponding to the subspace $\mathcal{B}_M^{(1)}$ of one-particle basis states localized in $M$: $\rho_M^{(1)}=(1/\mathcal{M})\mathrm{Tr}_M^{(1)}\ket{\widetilde{\Phi}}\bra{\widetilde{\Phi}}$, where $\mathcal{M}$ is a normalization constant such that $\mathrm{Tr}^{(1)}\rho_M^{(1)}=1$. The latter trace ($\mathrm{Tr}^{(1)}$) is meant within the complete one-particle basis $\mathcal{B}^{(1)}$ where $\rho_M^{(1)}$ is in general defined. Thus, we have
\begin{equation}\label{entanglement}
E_M(\Phi):= S(\rho_M^{(1)}) =-\sum_{i}\lambda_i\log_2 \lambda_i,
\end{equation}
where $S(\rho) = -\mathrm{Tr}(\rho\log_2 \rho)$ is the von Neumann entropy and $\lambda_i$ are the eigenvalues of $\rho_M^{(1)}$. 
Due to particle indistinguishability the amount of $E_M$ is in general expected to depend on $M$, that is on the localized region $M$ where the measurement is performed \cite{mintertreview,tichyFort}. When the particles are either spatially separated or in the same mode there cannot be any dependence on the localized mode where the measurement is done and we deem the entanglement obtained in this case as the ``intrinsic'', absolute entanglement of the system. 
We also notice that a wider scenario surfaces here regarding the entanglement determination for identical particles. In fact, after obtaining $\rho_M^{(1)}$ by locally tracing out one particle, the remained particle can be measured either again in the same localized mode $M$ or in a separated localized mode $M'$.

As an advancement with respect to the state-of-art, the described approach allows the quantification in complete generality of the physical entanglement of the two-particle system by directly applying the standard notion of partial trace to the assigned system state. This remarkable aspect makes our treatment very convenient and naturally generalizable to systems of many particles (to be addressed elsewhere), thus overcoming the drawbacks present in other specific name-labelled methods where redefinition of entanglement measures and witnesses are required \cite{ghirardi2004PRA,mintertreview,tichyFort,vogel2015PRA} (see appendixes \ref{Entderivation} and \ref{comparison} for some details on this point).

\subsection{Application} 
\begin{figure}[t!]
\begin{center}
{\includegraphics[width= 0.35\textwidth]{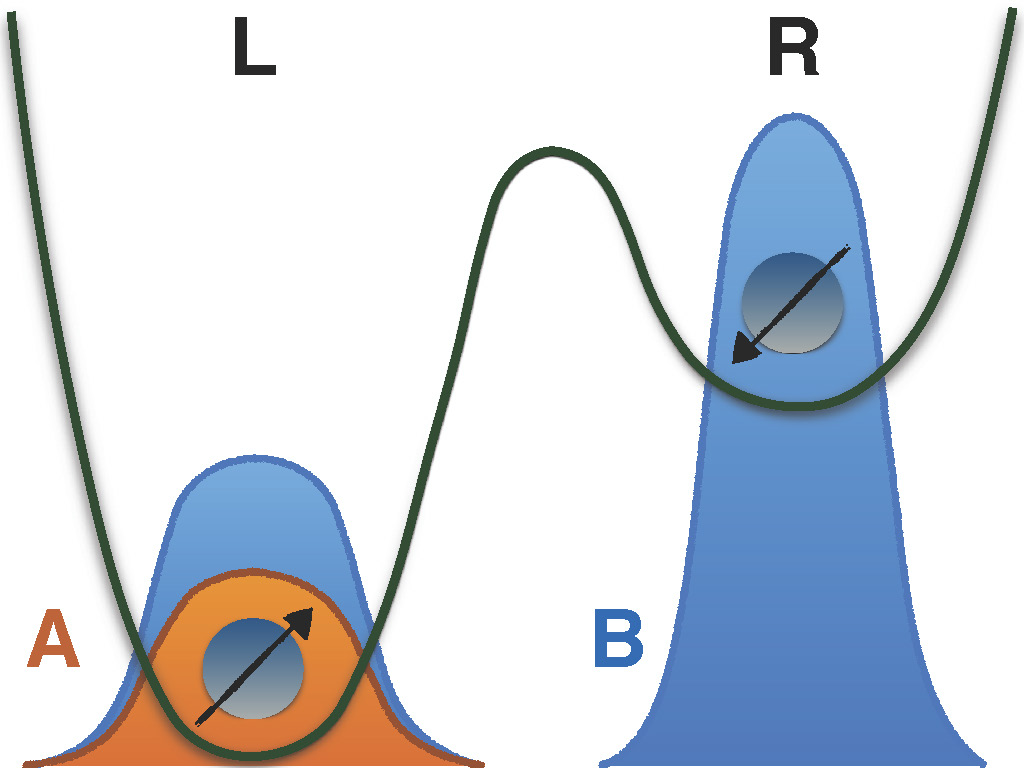}}
\end{center}
\caption{Asymmetric double-well. One particle is in the (orange) mode $\ket{A}$ equal to the localized ground state $\ket{L}$ of left well and one particle is in the (blue) mode $\ket{B}$ which is a combination of $\ket{L}$ and of the localized mode $\ket{R}$ of the right well, with $\scal{L}{R}=0$.}
\label{fig:doublewell}
\end{figure}

We now apply our method to the simplest case where entanglement may befall, that is a two-qubit system. We take each single-qubit state $\ket{\varphi}$ as $\ket{As}\equiv\ket{A}\ket{s}$, where $A$ indicates the spatial mode and $s=\uparrow,\downarrow$ two pseudospin internal states (e.g., components $\pm1/2$ of a spin-$1/2$ fermion, two energy levels of a boson, horizontal $H$ and vertical $V$ photon polarizations). 
We are interested in systems where spatial modes can overlap for an arbitrary extent. A simple system where this situation can happen is depicted in the asymmetric double-well configuration of Fig.~\ref{fig:doublewell}, where one particle is in the mode $\ket{A}=\ket{L}$ and one particle is in the mode $\ket{B}=\scal{L}{B}\ket{L}+\sqrt{1-|\scal{L}{B}|^2}\ \ket{R}$. This represents, for instance, a physical situation when at a given time the particle initially localized in $\ket{R}$ may slowly tunnel into $\ket{L}$, which may occur in Bose-Einstein condensates and quantum dots \cite{bloch2008RMP,petta2005Science,lundskog2014,tan2015PRL}.  
It is then convenient to study states whose structure make them entangled in the spatial separation scenario, as the Bell-like states.   
We thus choose two identical qubits prepared in the linear combination, valid for bosons, fermions and nonidentical particles, 
\begin{equation}\label{statePsi}
\ket{\Psi}=a\ket{L\uparrow, B\downarrow}+b e^{i\theta}\ket{L\downarrow, B\uparrow}, 
\end{equation}
where $a$ is positive real, $b=\sqrt{1-a^2}$ and $\scal{L}{B}\in[0,1]$. 

\begin{figure*}[tbph]
\begin{center}
{\includegraphics[height=4.8cm, width= 0.28\textwidth]{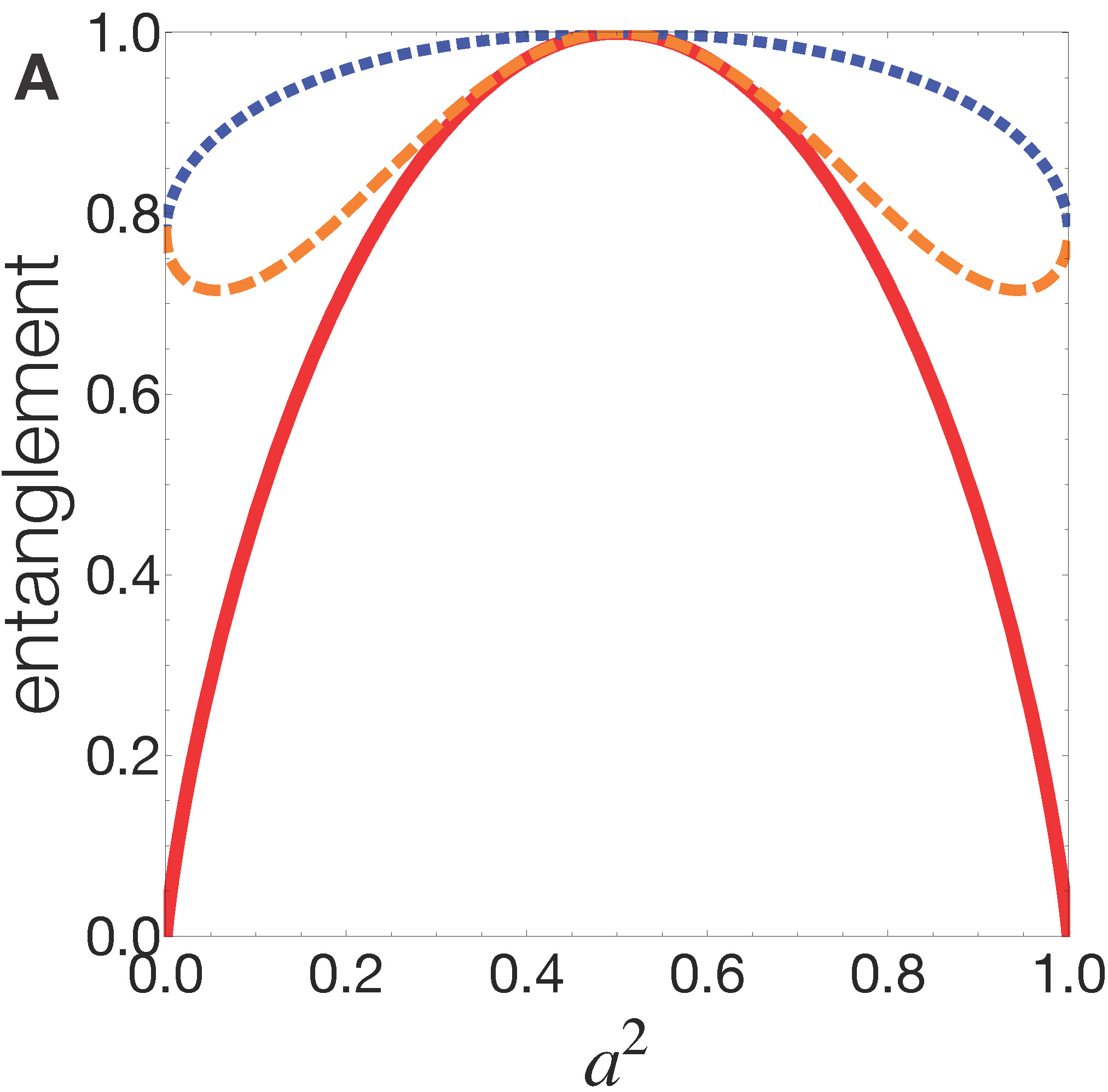}\hspace{0.1cm}
\includegraphics[width= 0.324\textwidth]{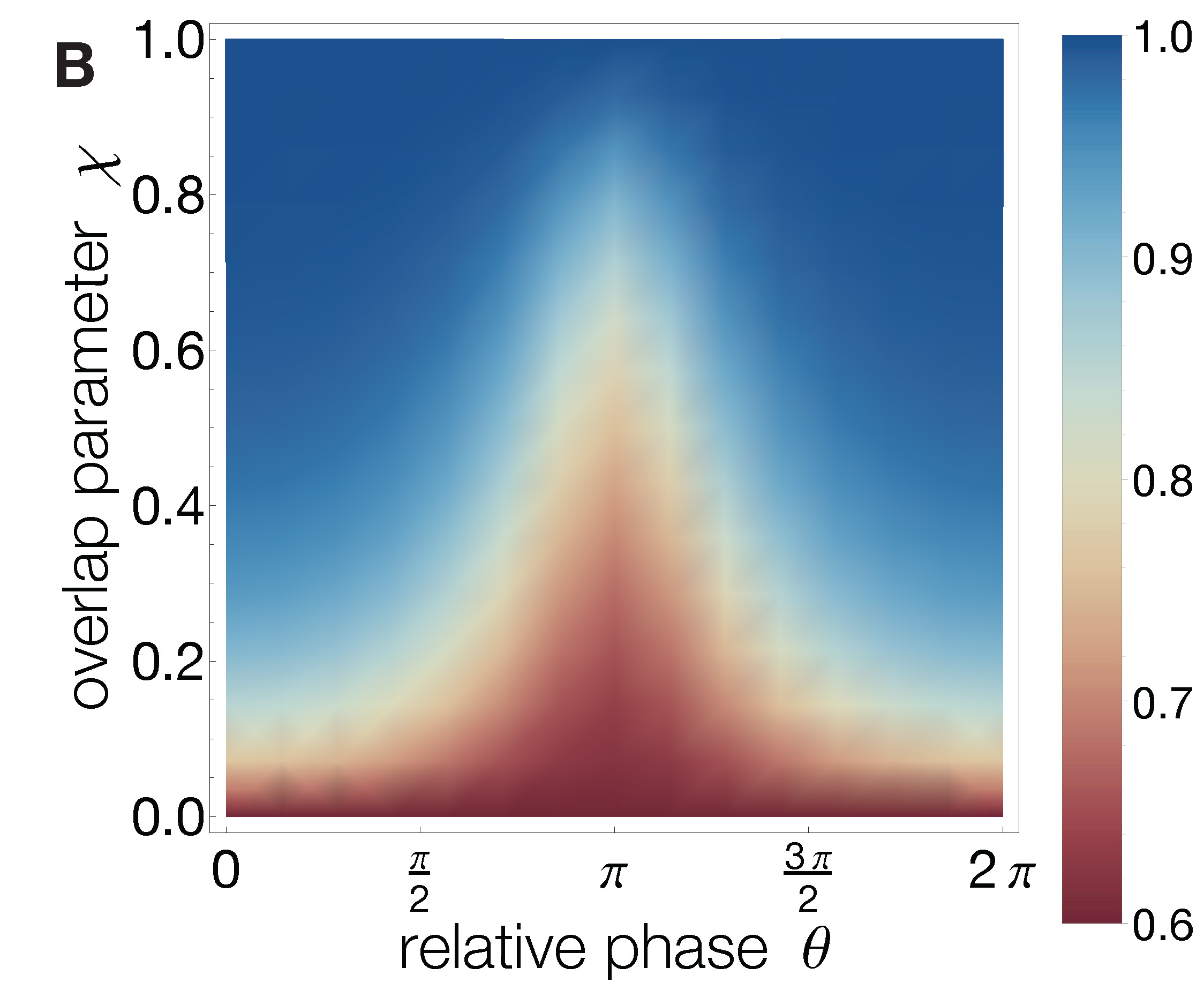}\hspace{0.1cm}
\includegraphics[width= 0.332\textwidth]{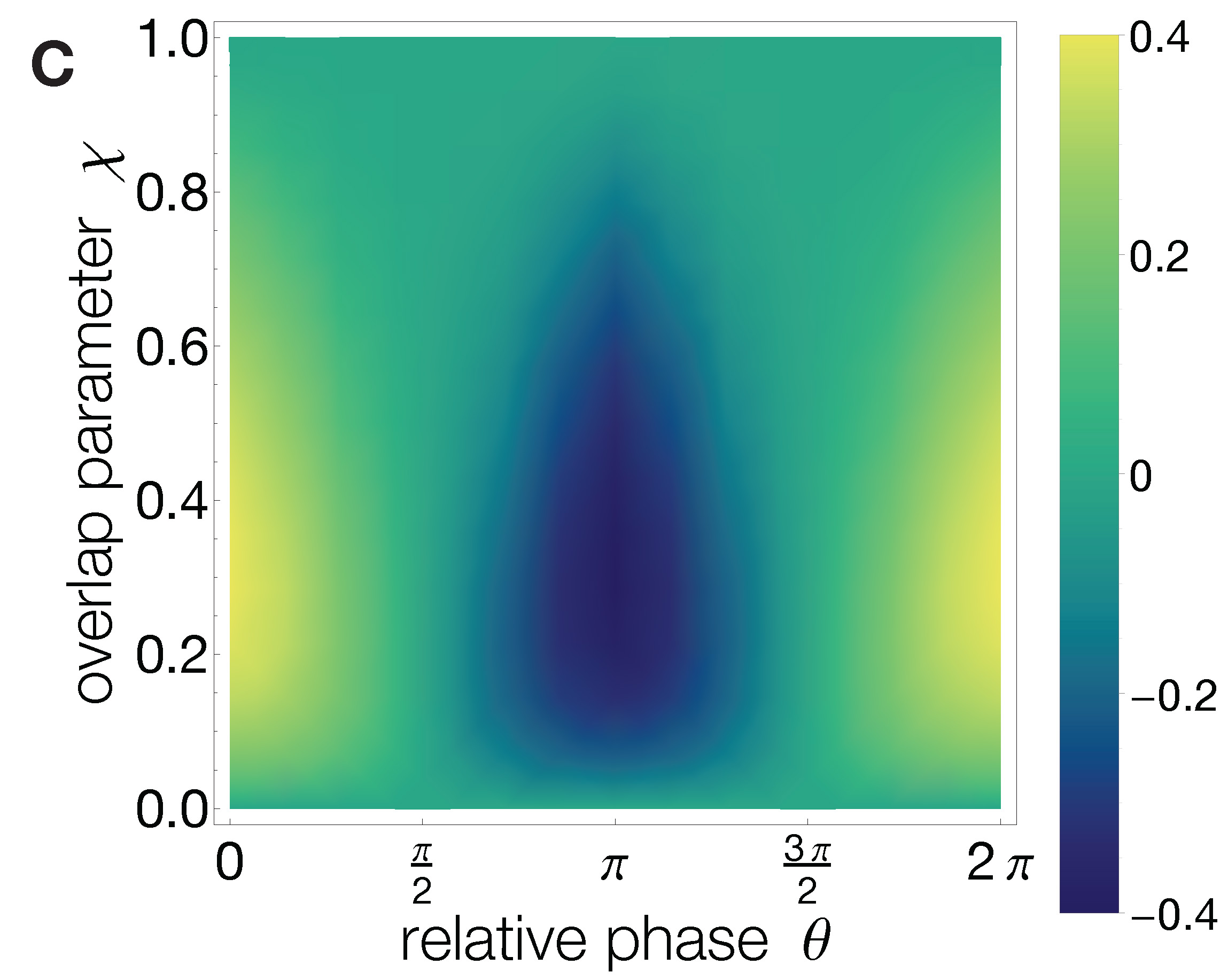}}
\end{center}
\caption{(\textbf{A}) Entanglement $E_L(\Psi)$ as a function of $a^2$ for $\theta=0$ and $\chi=0.3$ for bosons (blue dotted line) and fermions (orange dashed line), compared to the corresponding entanglement of nonidentical particles $E_\mathrm{ni}(\Psi)$ (red solid line). $E_L(\Psi)$ is always over the ``nonidentical particle fence'' delimited by $E_\mathrm{ni}(\Psi)$, collapsing to it when $\chi=0$. 
(\textbf{B}) Density plot of bosonic entanglement $E_L(\Psi)$, for $a=0.5$, as a function of both relative phase $\theta$ and overlap parameter $\chi$. The corresponding nonidentical particle entanglement, retrieved when $\chi=0$, is constantly equal to $E_\mathrm{ni}=0.634$. 
(\textbf{C}) Density plot of the difference between bosonic and fermionic entanglement, for $a=0.5$.}
\label{fig:identent}
\end{figure*}

Using Eq.~(\ref{partialtrace}), suitably normalized, with projective measurements onto the localized one-particle subspace $\mathcal{B}^{(1)}_L=\{\ket{L\uparrow},\ket{L\downarrow}\}$, it is straightforward to find the amount of entanglement $E_L(\Psi)$, which is given by Eq.~(\ref{entanglement}) with only two nonzero eigenvalues (see Supplemental Material)
\begin{equation}\label{Epartialoverlap}
\lambda_1=\frac{a^2+\chi(b^2+2\eta ab \cos\theta)}{1+\chi(1+4\eta ab \cos\theta)},  \quad 
\lambda_2=1-\lambda_1,
\end{equation}
where $\chi=|\scal{L}{B}|^2$ is the overlap parameter, which here coincides with the spatial mode fidelity. 
This result shows the quantitative manifestation of wave function overlap and particle statistics directly on the quantum entanglement of the system state, differently from other methods which require state tomography reconstruction at the level of separated detectors \cite{tichyFort}. In fact, our formalism permits the spatial overlap to explicitly prove responsible of a statistics-sensitive quantum interference phenomenon due to the phase in $\ket{\Psi}$ (see appendix \ref{Entderivation}). Entanglement $E_L(\Psi)$ is plotted in Fig.~\ref{fig:identent} to evidence its dependence on the nature of the particles. Relative phase $\theta$ and particle statistics $\eta$ have an effect for any $0<\chi<1$ (provided that $a\neq0, 1, 1/\sqrt{2}$). For states of the same form, entanglement is different for fermions and bosons: fermionic entanglement ($\eta=-1$) and bosonic one ($\eta=+1$) can be obtained one from the other by shifting $\theta$ of $\pi$.
When $\chi=0$ ($B=R$, orthogonal or spatially separated modes), one as expected gets $E_L(\Psi)=-a^2\log_2a^2-(1-a^2)\log_2(1-a^2)=E_\mathrm{ni}(\Psi)$ where $E_\mathrm{ni}(\Psi)$ is the entanglement entropy of nonidentical particles \cite{horodecki2009RMP,amico2008RMP}. 
This fact confirms that, under the condition of spatial separation, identical particles have the same entanglement of nonidentical particles \cite{facchiarxiv}. 
Moreover, for $\chi=0$ and $a=1$, $\ket{\Psi}$ becomes manifestly unentangled ($E=0$) as should be, a result to be contrasted with the entanglement entropy $S=\log_2 2=1$ obtained so far for such a state via partial trace within both first and second quantization approaches \cite{Li2001PRA,Paskauskas2001PRA,ghirardi2004PRA,mintertreview,balachandran2013PRL}. 
This fact made it necessary a rule to bypass this problem that however required a priori knowledge of the true quantum correlations in the state \cite{ghirardi2004PRA,mintertreview}.

When $a=1,0$, the state $\ket{\Psi}$ reduces to the elementary states $\ket{L\uparrow, B\downarrow}$, $\ket{L\downarrow, B\uparrow}$, respectively. In this case $E_L = \log_2 (1+\chi) - \frac{\chi}{1+\chi} \log_2 (\chi)$ for both bosons and fermions which ranges from 0 ($\chi=0$) to 1 ($\chi=1$), while for nonidentical particles one has $E_\mathrm{ni}=0$ since these states represent separable states and the one-particle reduced state is pure.

We now aim at answering unambiguously to the question whether two particles in the same site are entangled. We shall show that the answer depends on the configuration of the internal degrees of freedom (pseudospins). In particular, the state $\ket{L\uparrow, L\downarrow}$ ($\chi=1$) exhibits $E=1$: two identical particles in the same site with opposite pseudospins are maximally entangled. We adopt two argumentations, a conceptual and a definitional one, to further explain this result. On the one hand, it is physically intuitive to recognize that, if two identical particles in the same site with opposite spins are measured in their spins, it is impossible to distinguish which particle has a given spin because of the impossibility to individually address them. Indeed, the local spin measurement under this configuration is unavoidably ``nonlocal'' with respect to the particles, so implying the presence of entanglement in analogy to the entanglement induced by nonlocal measurements on uncorrelated spatially separated particles \cite{mintertreview}. However, the two-particle entanglement is here not induced but \emph{intrinsic} of the system due to its own spatial configuration. On the other hand, a two-particle state is maximally entangled if the two particles are perfectly correlated with respect to their spins when projecting (measuring) one particle along an arbitrary spin direction $\mathbf{u}$, as happens for Bell states of distinguishable particles \cite{horodecki2009RMP,amico2008RMP}. Our approach shows that this is the case for the two-particle elementary state $\ket{L\uparrow, L\downarrow}\equiv \ket{\uparrow, \downarrow}$ above (we omit the mode state $L$ for simplicity). By using Eq.~(\ref{reducedpurestate}), it is straightforward to see that 
\begin{eqnarray}\label{spincorrelation}
\eta=-1 &\Rightarrow& \scal{\uparrow_\mathbf{u}}{\uparrow,\downarrow}= \ket{\downarrow_\mathbf{u}}, \ \scal{\downarrow_\mathbf{u}}{\uparrow,\downarrow}=\ket{\uparrow_\mathbf{u}},\nonumber\\
\eta=+1 &\Rightarrow& \scal{\uparrow_\mathbf{u}}{\uparrow,\downarrow}= \ket{\downarrow_\mathbf{u'}}, \ \scal{\downarrow_\mathbf{u}}{\uparrow,\downarrow}=\ket{\uparrow_\mathbf{u'}},
\end{eqnarray}
where the versor $\mathbf{u'}$ is the reflection of $\mathbf{u}$ with respect to the $x$-$y$ plane (see appendix \ref{SpinProjections} for calculation details). Notice that the spin correlations found in Eq.~(\ref{spincorrelation}) for $\eta=\pm1$ (bosons and fermions) are exactly those occurring, respectively, for the Bell states of distinguishable particles $(\ket{\uparrow}\ket{\downarrow}\pm \ket{\downarrow}\ket{\uparrow})/\sqrt{2}$ by measuring a given (addressable) particle. The latter property provides theoretical support and prediction of the very recent experimental observations of spontaneous entanglement creation where such a state is generated by simply transporting two ultracold atoms of rubidium with opposite spins into the same optical tweezer \cite{kaufman2015Nature}. Contrarily, we observe that the entanglement of the state $\ket{L\uparrow, L\uparrow}\equiv \ket{\uparrow, \uparrow}$ (valid for bosons) is zero, as also verifiable by the fact that $ \scal{\uparrow_\mathbf{u}}{\uparrow,\uparrow}=\ket{\uparrow}$ independently of the first projective measurement: the state is uncorrelated (see appendix \ref{SpinProjections}). These results clarify the crucial role of both spatial mode and internal state configuration in assessing the entanglement of identical particles. 
  
It is now worth to observe that the entanglement of $\ket{\Psi}$ of Eq. (\ref{statePsi}) when determined by tracing out the particle spin in the localized mode $\ket{R}$ results $E_R(\Psi)= -a^2 \log_2 a^2 - (1-a^2) \log_2 (1-a^2)=E_\mathrm{ni}(\Psi)$ (see appendix \ref{Entderivation}). 
Putting these findings together shows how the entanglement fundamentally depends on the local measurement, a peculiar trait of identical particles which does not exist for distinguishable (or spatially separated) particles. 
In particular, the dependence of entanglement on the particle overlap appears only if both particles have nonzero probability of staying in the same localized mode where the measurement is done. The amount of entanglement naturally ceases to be dependent on the measurement localization only in the extremal situations when the particles are spatially separated or in the same mode, that is when we can speak about an intrinsic, absolute entanglement of the system. 
Identical particles prove more ``flexible'' than nonidentical ones in getting entangled: identical particle entanglement is never exceeded by the nonidentical particle counterpart.
 
We remark that a way to broadcast and exploit identical particle entanglement in the presence of spatial overlap is by extraction procedures, for instance, from Bose-Einstein condensates via tunneling \cite{bloch2008RMP} or from fermions (electrons) in quantum dots via photon emission \cite{lundskog2014}. The extraction of entanglement of overlapping identical particles has been so far investigated solely for the case of complete overlap based on particle labels \cite{cavalcanti2007PRB,plenio2014PRL}. This extraction can be also easily performed within our approach (see appendix \ref{extraction}). Our results directly indicate that the fundamental reason why entanglement can be extracted in this case is due to its intrinsic presence in the state.

\section{Discussion}
We have introduced a description of identical particles without particle labels which enables the treatment of their entanglement by means of notions usually adopted in entanglement theory for distinguishable particles, namely the von Neumann entropy of the reduced state after partial trace.
Our approach treats bosons, fermions and nonidentical particles on the same footing and is technically comparable to that used for nonidentical particles. This aspect significantly facilitates the quantitative study of entanglement under arbitrary conditions of particle wave function overlap and its generalization to many-particle systems for scalability. 

We have found that, contrarily to what happens for distinguishable particles, the amount of entanglement of identical particles does depend on local measurements. When two particles with opposite internal degrees of freedom (pseudospins) partially overlap in the same site where the projective measurements are done, entanglement is a function of their spatial overlap and an ``ordering'' emerges for different particle types. 
It is also of interest that the degree of entanglement associated to identical particles is never surpassed by the counterpart of nonidentical ones. This result indicates that identical particles may be more efficient than distinguishable ones for entanglement-based quantum information tasks. 
Summarizing, we have the following properties: (i) it is not possible, by a set of local measurements, to assess an absolute measurement-independent amount of entanglement for systems of identical particles unless the particles are spatially separated or in the same spatial mode; (ii) bringing identical particles into overlapping modes acts like an ``entangling gate'' whose effectiveness depends on the amount of overlap. Therefore, our approach clearly shows as a natural creation of maximally entangled states is possible just by moving two identical particles with opposite pseudospin into the same site, as recently confirmed in a recent experiment with ultracold atoms \cite{kaufman2015Nature}.
     
This study can in perspective lead to determine not only entanglement but also other kinds of quantum correlations \cite{Modi2012RMP,iemini2014} for systems of many identical particles, even in mixed states. Due to the importance of entanglement as a resource, its dynamical preservation is a crucial requirement \cite{xulofranco2013NatComms,adeline2014}: our analysis provides the groundwork for future investigations of this aspect.
Our results supply theoretical grounds in setting and interpreting experiments to use quantum correlations even in realistic scenarios where overlap of particles is important, as in condensed matter (Bose-Einstein condensates, spin chains, anyonic models) \cite{bloch2008RMP}, quantum dots \cite{petta2005Science,tan2015PRL}, superconducting circuits \cite{martinisfermions}, trapped ions \cite{solano2014}, photons in waveguides \cite{crespi2015} and biological molecular aggregates \cite{sarovarNatPhys,jesenkoNJP}. Our approach finally enables the extension of extraction procedures \cite{cavalcanti2007PRB,plenio2014PRL} for utilizing identical particle entanglement within the most general geometries of particle wave functions.


\appendix
\section{Particle exchange \label{exchange}}
One of the points that fundamentally differentiates the standard label-based approach from the present one is the role played by the exchange of particles. In fact, in the former the particle exchange operation is fundamental to construct a name label symmetrized state vector \cite{cohenbook}, here it is not. Nevertheless, physical exchange of two identical particles continues being a well-defined operation. After the exchange, the physical two-particle state is indistinguishable from before and the state vector acted by the exchange operator $\mathcal{P}$ remains the same up to a global phase factor: $\mathcal{P}\ket{\varphi_1,\varphi_2}=\pm\ket{\varphi_1,\varphi_2}$, the sign $+$ being associated to bosons and $-$ to fermions \cite{cohenbook}. Using then Eq.~(1), it follows $\mathcal{P}\ket{\varphi_1,\varphi_2}=\eta\ket{\varphi_1,\varphi_2}=\ket{\varphi_2,\varphi_1}$. Thus, the action of $\mathcal{P}$ on the two-particle state simply swaps the position of the one-particle states as they appear in the state. The definition of the two-particle probability amplitude of Eq.~(1) therefore directly encompasses the required particle spin-statistics symmetry, or symmetrization postulate \cite{cohenbook}. 

\section{Measurement-induced entanglement \label{measurement-induced}}
Consider two identical qubits in the state $\ket{L\uparrow, R\downarrow}$, where $L$, $R$ are two spatially separated orthogonal modes. This state is obtainable from the state $\ket{\Psi}=a\ket{L\uparrow, B\downarrow}+b\ket{L\downarrow, B\uparrow}$ considered in the main text for $a=1$ and $B=R$ (orthogonal modes). As discussed in the main text after Eq. (7), when the partial trace is performed by local projective measurement and the modes are spatially separated (orthogonal), the identical particles behave like nonidentical ones regarding their entanglement, that results to be $E=E_\mathrm{ni}=0$, as expected. 
Differently, we now perform the partial trace of Eq. (4) onto the nonlocal one-particle basis 
$\mathcal{B}^{(1)}_{LR}=\{\ket{\psi_1^\pm}=\ket{O_\pm\uparrow},\ket{\psi_2^\pm}=\ket{O_\pm\downarrow}\}$, where $\ket{O_\pm}=(\ket{L}\pm\ket{R})/\sqrt{2}$ are orthogonal nonlocal superpositions of the localized spatial modes $\ket{L}$, $\ket{R}$. Using Eqs.~(2) and (4), one easily obtains the ``measurement-induced" reduced density matrix 
$\rho_\mathrm{mi}^{(1)}=(\ket{L\uparrow}\bra{L\uparrow}+\ket{R\downarrow}\bra{R\downarrow})/2$, from which $E_\mathrm{mi}=S(\rho_\mathrm{mi}^{(1)})=1$: the  two qubits are maximally (measurement-induced) entangled. 
This fact confirms the peculiar property of entanglement of spatially separated identical particles to be sensitive to the nonlocal character of the measurement \cite{mintertreview}.

\section{Derivation of the identical particle entanglement \label{Entderivation}}
In this section we explicitly derive the identical particle entanglement given by Eqs.~(5) and (7) of the two-qubit state $\ket{\Psi}=a\ket{L\uparrow, B\downarrow}+be^{i\theta}\ket{L\downarrow, B\uparrow}$, where $a$ is positive real, $b=\sqrt{1-a^2}$ and $\scal{L}{B}\neq0$. Mode $\ket{B}$ is a general linear combination of $L$ and $R$ as: $\ket{B}=\scal{L}{B}\ket{L}+\sqrt{1-|\scal{L}{B}|^2}\ \ket{R}$. Notice that when $\scal{L}{B}=0$, $\ket{B}=\ket{R}$. The (mode-spin) one-particle basis is then $\mathcal{B}^{(1)}=\{\ket{L \uparrow},\ket{L \downarrow},\ket{R \uparrow},\ket{R \downarrow}\}$. Using this basis and the linearity property, the state $\ket{\Psi}$ can be written as 
\begin{eqnarray}\label{psi2}
\ket{\Psi}&=&\scal{L}{B}(a+\eta b e^{i\theta})\ket{L\uparrow, L\downarrow} \nonumber\\
&+&\sqrt{1-|\scal{L}{B}|^2}(a\ket{L\uparrow, R \downarrow} + b e^{i\theta}\ket{L\downarrow, R \uparrow}).
\end{eqnarray}  
In the state above, for partial overlap $0<\scal{L}{B}<1$, it is seen how both relative phase $\theta$ and particle statistics $\eta=\pm1$ are expected to play a role in the entanglement property.
In fact by using Eq. (2), suitably normalized, with projective measurements onto the localized one-particle subspace $\{\ket{L\uparrow},\ket{L\downarrow}\}$, we obtain the following one-qubit reduced density matrix expressed in the above basis $\mathcal{B}^{(1)}$
\begin{equation}\label{densitymatrix}
\rho_L^{(1)}=\frac{1}{\mathcal{M}}\left(\begin{array}{cccc} c_1& 0& c_4& 0\\ 0& c_1& 0& c_5\\ c_4^\ast& 0& c_2& 0\\ 0& c_5^\ast& 0& c_3
\end{array}\right),
\end{equation} 
where
\begin{eqnarray}\label{densitymatrixelements}
c_1&=&|\scal{L}{B}|^2(1+2\eta ab\cos\theta),\nonumber\\ 
c_2&=& b^2 (1-|\scal{L}{B}|^2),\ c_3=a^2  (1-|\scal{L}{B}|^2), \nonumber\\
c_4&=&\scal{L}{B}\sqrt{1-|\scal{L}{B}|^2}(a+\eta be^{i\theta})be^{-i\theta},\nonumber\\
c_5&=&\scal{L}{B}\sqrt{1-|\scal{L}{B}|^2}(a+\eta be^{i\theta})a,
\end{eqnarray} 
and $\mathcal{M}=2c_1+c_2+c_3$.
Diagonalizing this reduced density matrix we find only two nonzero eigenvalues $\lambda_1$, $\lambda_2$ as reported in Eq.~(7) of the main text, so that the identical particle entanglement is obtained by equation~(5) of the manuscript as $E_L(\Psi)=-\sum_{i=1,2}\lambda_i\log_2 \lambda_i$. 

Let us now determine the entanglement of the state $\ket{\Psi}$ of equation (8) by tracing out the particle in the other localized mode $\ket{R}$. Since there is no overlap between $\ket{A}$ and $\ket{R}$, one expects that the amount of overlap does not play any role in this case. In fact, it is easy to see that, using Eq. (2) suitably normalized with projective measurements onto the localized one-particle subspace $\{\ket{R\uparrow},\ket{R\downarrow}\}$, the one-particle reduced density matrix is localized in $L$ and reads
\begin{equation}\label{densitymatrixR}
\rho_R^{(1)}=a^2\ket{L\uparrow}\bra{L\uparrow} + b^2\ket{L\downarrow}\bra{L\downarrow},
\end{equation} 
which implies an entanglement entropy $E_R(\Psi)= -a^2 \log_2 a^2 - (1-a^2) \log_2 (1-a^2)$. A projection of the reduced density matrix $\rho_R^{(1)}$ on $L$ obviously does not change its form, being already localized in $L$, that is: $\rho_{RL}^{(1)}=\frac{\Pi_L\rho_R^{(1)}\Pi_L}{\mathrm{Tr}(\Pi_L\rho_L^{(1)}\Pi_L)}=\rho_R^{(1)}$, where $\Pi_L=\ket{L}\bra{L}$.

We also notice that a wider scenario surfaces here regarding the entanglement determination for identical particles. In fact, after obtaining $\rho_M^{(1)}$ by localized partial trace, the remained particle can be measured either again in the same localized mode $M$ or in a separated localized mode $M'$.
We can also choose to study the entanglement when the first partial trace is made on $L$ and the second trace is made on $R$. By projecting $\rho_L^{(1)}$ of Eq.~(9) on $\ket{R}$ by means of the projection operator $\Pi_R=\ket{R}\bra{R}$ and renormalizing we get
\begin{equation}\label{densitymatrixLR}
\rho_{LR}^{(1)}=\frac{\Pi_R\rho_L^{(1)}\Pi_R}{\mathrm{Tr}(\Pi_R\rho_L^{(1)}\Pi_R)}=a^2\ket{R\downarrow}\bra{R\downarrow} + b^2\ket{R\uparrow}\bra{R\uparrow},
\end{equation} 
which coincides with $\rho_{RL}^{(1)}$ and gives an entanglement entropy $E_{LR}(\Psi)=E_R(\Psi)$.

We finally calculate the entanglement when the first partial trace is made on $L$ and the second trace is made again on $L$. By projecting $\rho_L^{(1)}$ on $\ket{L}$ by means of the projection operator $\Pi_L=\ket{L}\bra{L}$ and renormalizing we get
\begin{equation}\label{densitymatrixLL}
\rho_{LL}^{(1)}=\frac{\Pi_L\rho_L^{(1)}\Pi_L}{\mathrm{Tr}(\Pi_L\rho_L^{(1)}\Pi_L)}=\frac{1}{2}(\ket{L\downarrow}\bra{L\downarrow} + 
\ket{L\uparrow}\bra{L\uparrow}),
\end{equation} 
which provides an entanglement entropy $E_{LL}(\Psi)=1$. 

These results show how identical particle entanglement quantitatively depends on the way the local measurement is performed. We notice that such a behavior already surfaced in the entanglement measured by the so-called ``detection-level concurrence'' $C_d$ \cite{tichyFort}, which also includes the effects of spatial overlap and particle statistics. However, this new notion of entanglement measure $C_d$, introduced specifically for identical particles, presents drawbacks regarding the case of particles in the same site and generalizations to many particles (see discussion in appendix \ref{comparison}).

\section{Entanglement extraction \label{extraction}}
We show here that identical particle entanglement extraction can be obtained within our approach, without resorting to particle name labels. For simplicity, we consider two identical qubits in the state $\ket{\Phi'^{(2)}}=\ket{L\uparrow,L\downarrow}$. This state is obtainable from $\ket{\Psi}=a\ket{L\uparrow, B\downarrow}+b\ket{L\downarrow, B\uparrow}$ considered in the main text for $\chi=|\scal{L}{B}|^2=1$  ($B=L$). We already know from the manuscript that this state is maximally entangled. This identical particle entanglement is unexploitable by local operations and classical communication (LOCC) because the single particles (subsystems) cannot be individually addressed.
We then consider a unitary spin-insensitive splitting transformation (tunneling) $\ket{L,s}\rightarrow r\ket{C,s}+t\ket{D,s}$, where $s=\uparrow,\downarrow$ and $|r|^2+|t|^2=1$, that transfers a particle from the localized mode $L$ to the two distinct modes $C$, $D$ with amplitudes $r$, $t$, respectively. Such a transformation represents a beam splitter transformation from optics that, for instance, is equivalent to a tunneling operation for Bose-Einstein condensates where particles can leak from mode $A$ into neighboring modes $C$, $D$ \cite{plenio2014PRL}. 
Applying the splitting transformation to the input state $\ket{\Phi'^{(2)}}=\ket{L\uparrow,L\downarrow}$ we immediately obtain the normalized output state
\begin{equation}\label{outputextraction}
\ket{\Phi_\mathrm{out}^{(2)}}=r^2 \ket{\Phi_{C}^{(2)}} +\sqrt{2} r t \ket{\Phi_{CD}^{(2)}}+t^2 \ket{\Phi_{D}^{(2)}}
\end{equation} 
where
$\ket{\Phi_{C}^{(2)}}=\ket{C\uparrow,C\downarrow}$,  $\ket{\Phi_{D}^{(2)}}=\ket{D\uparrow,D\downarrow}$ and
$\ket{\Phi_{CD}^{(2)}}= (\ket{C\uparrow,D\downarrow}+\ket{C\downarrow,D\uparrow})/\sqrt{2}$.
States $\ket{\Phi_{C}^{(2)}}$ and $\ket{\Phi_{D}^{(2)}}$ are exactly of the same type of the input state, with all the particles being in a single mode: these two states thus have the same entanglement of the input state but this entanglement remains unexploitable by LOCC. The state $\ket{\Phi_{CD}^{(2)}}$ instead is a maximal linear combination of two identical particles in two orthogonal distinct modes (like a Bell state). We know from the main text that this state is maximally entangled. This entanglement is now an exploitable resource within the LOCC framework because it is (probabilistically) established between two particles in accessible distinguishable (spatially separated) modes. 

We remark that the entanglement of three identical bosons in the same mode $\ket{\Phi'^{(3)}}=\frac{1}{\sqrt{2}}\ket{L\downarrow,L\downarrow,L\uparrow}$, which is exactly the same state considered in Ref.~\cite{plenio2014PRL}, can be straightforwardly extracted within our approach by following analogous calculations. In Ref.~\cite{plenio2014PRL} the extraction procedure required particle name labels, but we demonstrate that particle names, although useful in this specific case, are unneeded to perform this operation.

\section{Comparing our particle-based approach with the usual name-labelled one \label{comparison}}
In our particle-based approach with no labels, the elementary state of two independent identical particles having orthonormal states $\varphi_1$ and $\varphi_2$ is simply $\ket{\Phi^{(2)}}=\ket{\varphi_1,\varphi_2}$. In the standard first quantization approach, where name labels are assigned to the particles, the same state is given by $\ket{\psi(1,2)}=(\ket{\varphi_1^{(1)}}\otimes\ket{\varphi_2^{(2)}}\pm \ket{\varphi_2^{(1)}}\otimes\ket{\varphi_1^{(2)}})/\sqrt{2}$, where $+$ is for bosons (symmetrized state) and $-$ for fermions (antisymmetrized state), respectively \cite{cohenbook,feynmanquantum}. We stress that our state $\ket{\Phi^{(2)}}$ is meant as a whole object and does not insinuate at all the presence of name labels for the particles. Differently, the state $\ket{\psi(1,2)}$, which has the structure of an entangled state with respect to the labels (1, 2), it is meant as a product state $\ket{\varphi_1}\otimes\ket{\varphi_2}$ of the two particles since it is obtainable by symmetrizing (antisymmetrizing) just this state \cite{ghirardi2004PRA}. 
To stress this difference, we notice that in our framework an entangled state of spatially separated identical particles has always the form $a\ket{\varphi_1,\varphi_2} + b\ket{\varphi'_1,\varphi'_2}$, exactly as happens for distinguishable particles. Such a state in the standard description with labels has, instead, four ket states.

We then remark that the one-particle projective measurement of Eq. (\ref{reducedpurestate}) and the partial trace operation of Eq. (\ref{partialtrace}) defined in our approach are physical operations performed on the two-particle system, giving the physical reduced state of the system, where it is not contemplated the addressing of an individual particle. At variance, a one-particle partial trace performed on a particle with a given name does not correspond to a physical trace. This is what happens in the name-labelled approaches for one-particle partial trace operations which are mere mathematical tool \cite{ghirardi2004PRA,mintertreview}. The latter fact gives rise to problems and contradictions between fermions and bosons in taking the von Neumann entropy of the reduced state as a faithful entanglement measure (see, for instance, discussions at pages 5-7 of Ref. \cite{ghirardi2004PRA}). Our \emph{physical} partial trace operation conceptually differs from this one, never suffers these issues and always provides the physical entanglement of the system. 

We now consider another important point. A standard criterion employed by the name-labelled approaches to witness entanglement is that, if a two-particle state is obtainable after symmetrization (antisymmetrization) with respect to the labels of a product state of orthonormal states, then it must be considered unentangled \cite{ghirardi2004PRA,mintertreview}. This assumption immediately excludes the possibility that an elementary state of two identical particles in orthogonal states can be entangled. In fact, in this case one retrieves the state $\ket{\psi(1,2)}$ which is unentangled according to this criterion. 
The consequence of this fact is that two identical particles located in the same spatial mode $A$ (complete overlap) with orthogonal internal states (e.g., opposite spins) are unentangled. We notice that this assumption and the same conclusion are also contained in the so-called ``detection-level concurrence'' \cite{tichyFort}, which quantifies the entanglement of the two-particle state (expressed in terms of name labels) constructed by state tomography at spatially separated detectors where the particles are measured. But these results contrast with the findings obtained by entanglement extraction procedures \cite{cavalcanti2007PRB,plenio2014PRL} and recent experimental observations \cite{kaufman2015Nature,veldhorst2015Nature} which give a strong evidence of entanglement for identical particles with opposite spins in the same site. Our approach, where elementary identical particle states are assumed as a whole entity as should physically be, provides theoretical support to these observations. In general: an elementary two-particle state (e.g., $\ket{A\uparrow, B\downarrow}$) can be entangled even if the one-particle states are orthogonal ($\scal{A\uparrow}{B\downarrow}=0$) provided that the spatial modes overlap ($\scal{A}{B}\neq 0$).

It is also worth noticing that the mentioned detection-level concurrence \cite{tichyFort}, besides missing the intrinsic entanglement present between two identical particles in the same site, is not suitable for treating entanglement in many-particle systems since it would require the prohibitive construction by quantum state tomography of the overall state at the level of spatially separated detectors. The power of our approach becomes evident also in this respect: being based on the standard notion of (physical) partial trace on particles of the system, it indeed enables a natural extension to many-particle systems in analogy to the case of distinguishable particles.

\section{Spin correlations for two identical particles in the same site \label{SpinProjections}}
Let us take two identical particles with opposite spins in the same spatial mode, described by the state $\ket{L\uparrow,L\downarrow}\equiv \ket{\uparrow,\downarrow}$, and consider an arbitrary direction for the spin individuated by the versor $\mathbf{u}=(\sin\theta \cos\varphi, \sin\theta \sin\varphi, \cos\theta)$, where $\theta$ and $\varphi$ are the polar and azimuthal angle, respectively. We want to find what happens to the remaining particle of the state above after projecting one particle along the spins $\ket{\uparrow_\mathbf{u}}=\cos(\theta/2)\ket{\uparrow}+e^{\mathrm{i}\varphi}\sin(\theta/2)\ket{\downarrow}$ and 
$\ket{\downarrow_\mathbf{u}}=-e^{-\mathrm{i}\varphi}\sin(\theta/2)\ket{\uparrow}+\cos(\theta/2)\ket{\downarrow}$ \cite{cohenbook}. 
By using Eq.~(\ref{reducedpurestate}) we easily find
\begin{eqnarray}\label{spincorrelation1}
\scal{\uparrow_\mathbf{u}}{\uparrow,\downarrow}&=& \cos(\theta/2)\ket{\downarrow} +\eta e^{-\mathrm{i}\varphi} \sin(\theta/2)\ket{\uparrow},
\nonumber\\
\scal{\downarrow_\mathbf{u}}{\uparrow,\downarrow}&=&-e^{\mathrm{i}\varphi}\sin(\theta/2)\ket{\downarrow}+\eta\cos(\theta/2)\ket{\uparrow}.
\end{eqnarray}
It is immediately seen that, after each of the one-particle measurements above, the other particle is left in, respectively: (i) for fermions, $\eta=-1$, $\ket{\downarrow_\mathbf{u}}$, $\ket{\uparrow_\mathbf{u}}$; (ii) for bosons, $\eta=+1$, $\ket{\downarrow_\mathbf{u'}}$, $\ket{\uparrow_\mathbf{u'}}$, where $\mathbf{u'}$ is the versor defined by the angles $\theta'=\pi-\theta$ and $\varphi'=\varphi$. Notice that the same results are obtained for distinguishable particles in the Bell states $(\ket{\uparrow}\ket{\downarrow}\pm \ket{\downarrow}\ket{\uparrow})/\sqrt{2}$ by projecting the first particle, for instance, along the spin directions above. 

On the contrary, let us see what happens for a spin projective measurement on the (unnormalized) state $\ket{L\uparrow,L\uparrow}\equiv \ket{\uparrow,\uparrow}$. We immediately see that
\begin{equation}\label{spincorrelation2}
\scal{\uparrow_\mathbf{u}}{\uparrow,\uparrow}= 2\cos(\theta/2)\ket{\uparrow},
\end{equation}
which, apart a normalization constant, gives the one-particle state $\ket{\uparrow}$ independently of the first measurement. The state $\ket{L\uparrow,L\uparrow}$ is thus unentangled (separable or uncorrelated).

The above results testify the perfect spin correlations between the two identical particles in the same site, which behave exactly like being maximally entangled. No particle has a definite state but the two particles are a whole object in a perfectly correlated (non-separable) state.

\end{document}